\documentclass[10pt]{article}
\usepackage[letterpaper,hmargin=1.4in,vmargin=1.4in]{geometry} 
\usepackage{xspace,url,multirow}
\usepackage{latexsym}
\usepackage{amssymb}
\usepackage{amsmath}
\usepackage{graphicx,wrapfig}
\usepackage[ruled,lined,linesnumbered]{algorithm2e}
\usepackage[belowskip=-3mm,aboveskip=2pt]{caption}
\usepackage[colorlinks=false,bookmarks=false,hyperfootnotes=false]{hyperref}

\newtheorem{lem}{Lemma}

\newtheorem{defn}{Definition}
\newtheorem{thm}{Theorem}
\newcommand{\etal}{et~al.}
\def \proof {\par \noindent {\bf Proof.}\hskip 5pt}
\def \endproof {\hfill $\Box$ \smallskip}
\def \safe {\text{safe}}
\def \bound{\kappa}
\def \r{\widetilde r}
\def \toadd#1 {\noindent \texttt{To add: #1} \\ \noindent}

{\makeatletter
 \gdef\xxxmark{%
   \expandafter\ifx\csname @mpargs\endcsname\relax 
     \expandafter\ifx\csname @captype\endcsname\relax 
       \marginpar{xxx}
     \else
       xxx 
     \fi
   \else
     xxx 
   \fi}
 \gdef\xxx{\@ifnextchar[\xxx@lab\xxx@nolab}
 \long\gdef\xxx@lab[#1]#2{\textbf{[\xxxmark #2 ---{\sc #1}]}}
 \long\gdef\xxx@nolab#1{\textbf{[\xxxmark #1]}}
}

\author{Immanuel Halupczok\footnote{{ Institut f\"ur Mathematische Logik und Grundlagenforschung, Universit\"at M\"unster,
Germany.
{ \texttt{[ihalu\_01|andre.schulz]@uni-muenster.de}}
}} \and  Andr\'e Schulz\footnotemark[1]}

\title{Pinning Balloons with Perfect Angles and Optimal Area}

\begin{document}

\maketitle

\begin{abstract}
We study the problem of arranging a set of $n$ disks with prescribed radii on $n$ rays emanating from the origin  such that two neighboring rays are separated by an angle of $2\pi/n$. 
The center of the disks have to lie on the rays, and no two disk centers are allowed to lie on the same ray.
We require that the disks have disjoint interiors, and that for every ray the segment between the origin and the boundary of its associated disk avoids the interior of the disks. 
Let $\r$ be the sum of the disk radii.
We introduce a greedy strategy that constructs such a disk arrangement that can be covered with a disk  centered at the origin whose radius is at most $2\r$, which is best possible.
The greedy strategy  needs $O(n)$ arithmetic operations.

As an application of our result we present an algorithm for embedding unordered trees with straight lines and perfect angular resolution such that it can be covered with a disk of radius $n^{3.0367}$, while having no edge of  length smaller than 1. The tree drawing algorithm is an enhancement of  a recent result by Duncan et~al.~[Symp.~of Graph Drawing, 2010] that exploits the heavy-edge tree decomposition technique to construct a drawing of the tree that can be covered with a disk of radius  $2 n^4$. 
\end{abstract}

\section{Introduction}

When a graph is drawn in the plane, the vertices are usually represented as small dots. From a theoretical point of view a vertex is realized as a point, hence as an object without volume. In many applications, however, it makes sense to draw the vertices as disks with volume. The radii of the vertices can enhance the drawing by visualizing associated vertex weights~\cite{BGR03,DEKW06}. This idea finds also applications in so-called \emph{bubble drawings}~\cite{GADM04}, and \emph{balloon drawings}~\cite{LY07,LYPF11}.

Two important quality measures for  aesthetically pleasant drawings are the \emph{area} of a drawing and its \emph{angular resolution}. The area of a drawing denotes the area of the smallest disk that covers the drawing with no edge lengths smaller than 1. The angular resolution denotes the minimum angle between two neighboring edges emanating at a vertex. Unfortunately, drawings of planar graphs with bounded angular resolution require exponential area~\cite{MP94}. On the other hand, by a recent result of Duncan~\etal~\cite{DEGKN10}, it is possible to draw any unordered tree as plane straight-line graph with \emph{perfect angular resolution}, that is the edges incident to a vertex $v$ are separated by an angle of at least $2\pi /\text{degree}(v)$, and polynomial area. In the same paper it was observed that an ordered tree drawn with perfect angular resolution requires exponential area. Surprisingly, even ordered trees can be drawn in polynomial area with perfect angular resolution when the edges are drawn as circular arcs~\cite{DEGKN10}. 

The following sub-problem appears naturally in tree drawing algorithms. Suppose we have drawings of all subtrees of the children of the root.
How can we group the subtrees around the root, such that the final drawing is densely packed? 
Often one assumes that every subtree  lies exclusively in some region, say a disk. Hence, at its core, a tree drawing algorithm has to arrange disjoint disks ``nicely'' around a new vertex. Furthermore this task is also a fundamental base case for bubble drawing algorithms or for algorithms that realize vertices as large disks. 
In the paper we show how to layout the balloons with perfect angular resolution and optimal area.

More formally, 
let $\mathcal{B}=\{B_1, B_2,\dotsc, B_n\}$ be a
 set of  $n$ disks. To distinguish the disks $B_i$ from other disks we call them \emph{balloons}. The balloon $B_i$ has radius $r_i$, and the balloons are sorted in increasing order of their radii. We are interested in layouts, in which the  balloons of $\mathcal{B}$  have disjoint interiors and are evenly angularly spaced. In particular, we draw for every balloon a \emph{spoke}, that is a line segment from the origin to the balloon center. The spokes have to avoid the interior of the other balloons and two neighboring spokes are separated by an angle of $2\pi/n$. 
Furthermore the drawing should require only small area. We measure the area of the balloon layout by the radius of the smallest disk that is  centered at the origin and  covers all balloons.
 
\paragraph{Results.}
We show how to locate the balloons with perfect angular resolution such that the drawing can be covered with a disk of radius $2\r$, for  $\r$ being the sum of the radii. This is clearly the best possible result in the worst case, since when $|\mathcal{B}|=1$, the area of the best balloon layout is clearly   $2 r_1$.
We also study a modified version of the balloon layout problem that finds application in a tree drawing algorithm. Here, one and two spokes may remain without balloon, but the angle between the two unused spokes has to be at least $2\pi/3$. In this setting we obtain a balloon drawing that  can be covered with a disk of radius $(1+\sqrt{2-2/\sqrt{5}}) \r \approx 2.0514 \r$. The induced  algorithm draws unordered trees with perfect angular resolution and with area smaller  than $n^{3.0367}$.
 
\paragraph{Related work.}
Without explicitly stated, Duncan~\etal~\cite{DEGKN10} studied the balloon layout  problem (with one or two unused spokes) as part of their drawing algorithm for unordered trees and obtained a bound of $4 \r$ for the area.  The induced tree drawing algorithm produces drawings with area smaller than $2n^4$. For the special case of orthogonal straight-line drawings of ternary trees (they automatically guarantee perfect angular resolution) Frati~\cite{F07} provided an algorithm whose drawings require $O(n^{1.6131})$ area; the drawing of the complete ternary tree requires $O(n^{1.262})$ area.  Bachmaier~\etal~obtained a drawing of the complete 6-regular tree with perfect angular resolution with area $O(n^{1.37})$~\cite{BBB+08}. In contrast to our setting the so-called balloon drawings~\cite{LY07,LYPF11} place all balloons at the same distance. Also related are the (non-planar) \emph{ringed circular layouts}~\cite{TM02}. Without the perfect angular resolution constraint trees can be drawn with area $O(n\log n)$~\cite{CBP92}.


\paragraph{Conventions.}
We normalize the radii of the balloons such that they sum up to 1.  In intermediate stages of the drawing algorithm a spoke may be without a balloon. In this case we consider the spoke as a ray emanating from the origin that fulfills the angular resolution constraint. When we say that ``we place balloon $B$ on $s$ at distance $x$'' we mean that the balloon $B$ is placed on a spoke $s$ (that had no associated balloon yet) such that its center  lies on $s$ at Euclidean distance $x$ from the origin. In the remainder of the paper all disks covering the balloons are considered as centered at the origin. 

\section{The greedy strategy}\label{sec:basic}
In the following section we introduce the greedy strategy for placing $\mathcal{B}$ with perfect angles. To keep things simple we assume for now that the number of balloons $n$ is a power of two. The general case is discussed later.

\begin{figure}[htb]
\begin{center}
\begin{tabular}{cp{1cm}c}
  \includegraphics[width=.33\columnwidth]{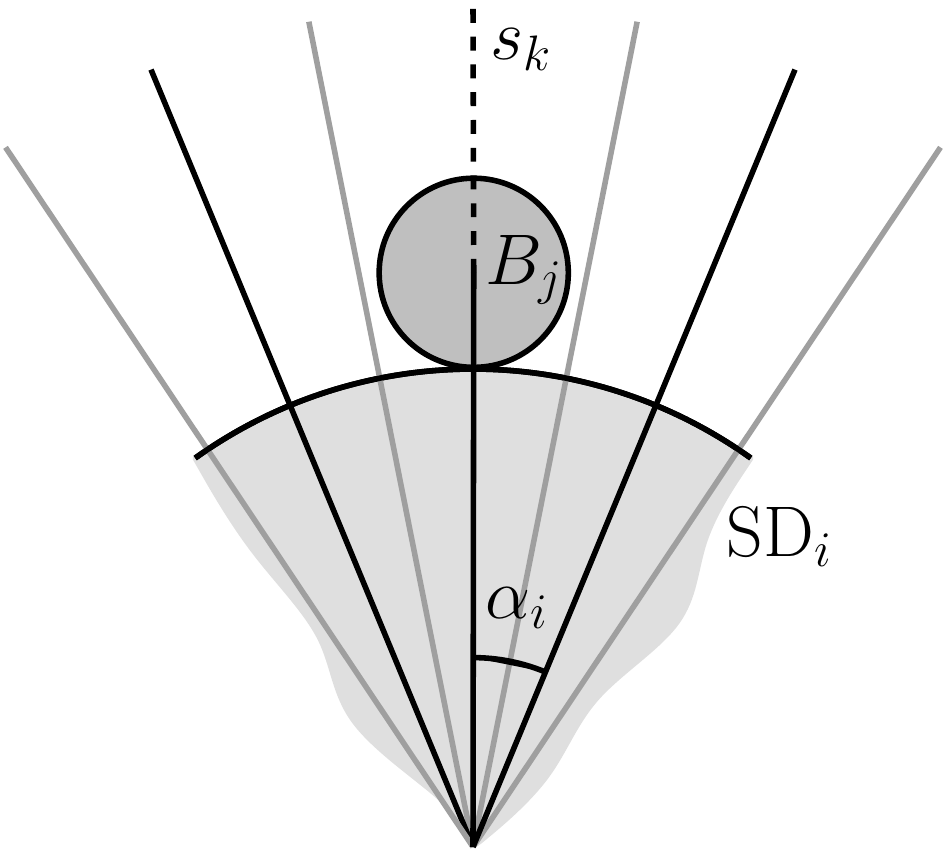} & &
  \includegraphics[width=.33\columnwidth]{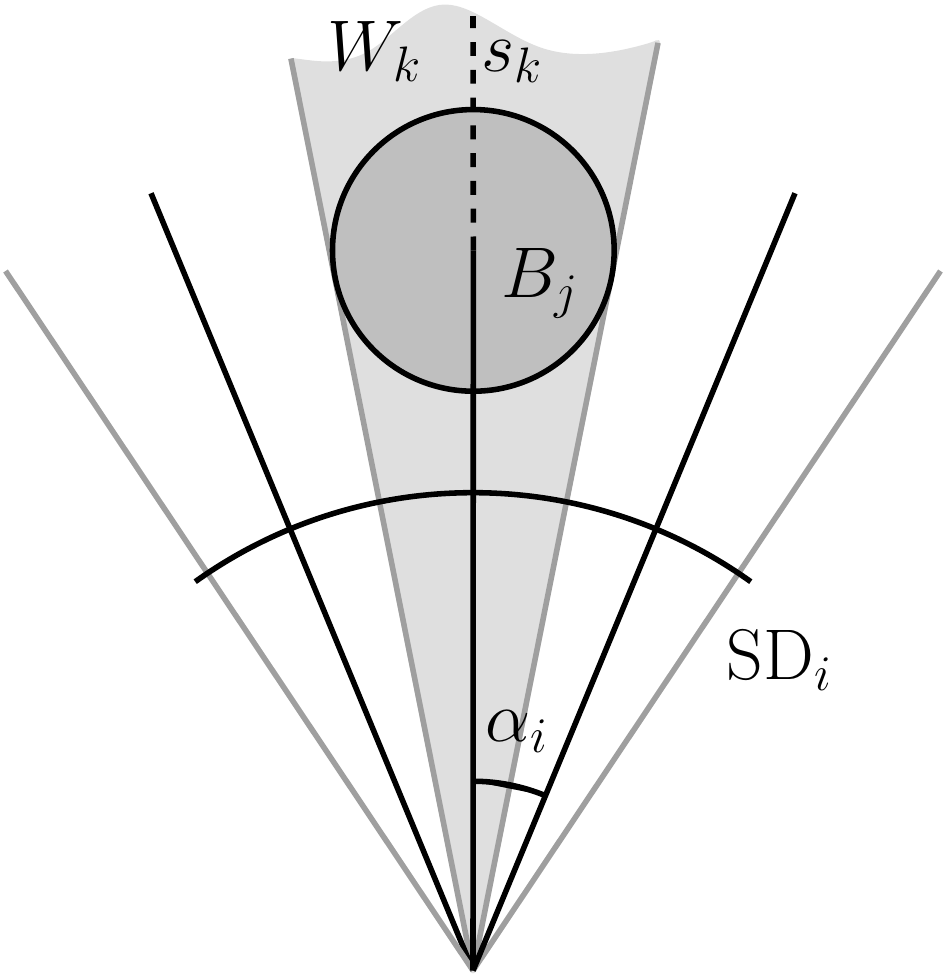} 
  \\
\small{(a)} & & \small{(b)} 
\end{tabular}
\caption{In a contact situation (a) we place $B_i$ such that it touches $\text{SD}_i$. In contrast, in a wedge situation (b), we place $B_i$ such that it touches the boundary of $W_k$ (when it is placed on $s_k$).}
  \label{fig:situation}
\end{center}
\end{figure}

We place the balloons in  increasing order of their radii. Thus we start with the smallest balloon and end with the largest balloon. The placement of the balloons is carried out in \emph{rounds}. In every round we locate half of the balloons that have not been placed yet. Thus, we ``consume'' a certain number of spokes in each round. Let $S$ be the list of spokes that are  available in the beginning of a round  in cyclic order. In every round we select every other spoke as a spoke on which a balloon is placed in the current round. This ensures that consecutive spokes that receive a balloon in round $i$  are separated by an angle of $\alpha_i:=2^{i+1}\pi/n$. For every round we define the \emph{safe disk} $\text{SD}_i$ centered at the origin with radius $\safe_i$. The safe disk is the smallest disk covering all balloons that were placed in previous rounds.
In round $i$ we place all balloons such that they   
 avoid the interior of the safe disk $\text{SD}_i$. Thus, the best we can hope for is to place the balloons such that they touch $\text{SD}_i$. Whenever this is possible we speak of a \emph{contact situation}, depicted in Figure~\ref{fig:situation}(a). The safe disks ensure that balloons placed in the current round will not intersect the interior of the balloons that were placed in previous rounds. However, we have to  guarantee that balloons placed in the same round will also not interfere with the remaining spokes. Suppose that $B_j$ is assigned to the spoke $s_k$. We enforce $B_j$ to lie inside a wedge with opening angle $\alpha_i$ centered at $s_k$. This wedge is named $W_k$. Since the spokes that are used in round $i$ are separated by $\alpha_i$, the wedges of round $i$ have disjoint interiors. Whenever a balloon touches the boundary of its associated wedge we speak of a \emph{wedge situation}, as shown in Figure~\ref{fig:situation}(b). 

The greedy strategy tries first to place $B_j$ at its spoke $s_k$, such that it touches $\text{SD}_i$. If this would imply that $B_j$ is not contained inside $W_k$, we move the center of $B_j$ on $s_k$ away from the origin, until $B_j$ touches the boundary of $W_k$. In case a wedge situation occurs, we can compute the location of the center of $B_j$ with help of the following lemma, which was also proven in a slightly different form by Duncan~\etal~\cite{DEGKN10}.
\begin{lem}\label{lem:wedgedisk}
Let $W$ be a wedge with opening angle $\varphi$ centered at a spoke $s$. Further let $B$ be a balloon with radius $r$ that is placed such that (1) its center lies on $s$, and (2) it touches the boundary of $W$. Then $B$ is contained inside a disk centered at the origin with radius
\[\frac{1+\sin{(\varphi /2)}}{\sin{(\varphi /2)}} \cdot r .\]
\end{lem}
\proof
Let $T$ be the point where $B$ touches the boundary of $W$ and $C$ the center of $B$. The triangle spanned by the origin, $T$, and $C$ has a right angle at $T$ and an angle $\alpha/2$ at the origin. Therefore, $C$ has distance $r/\sin{(\alpha/2)}$ from the origin. To cover $B$ we add $r$ to  the radius of the disk that touches $C$. 
The resulting radius equals $r+r/\sin{(\alpha/2)}$.
\endproof
%
In the remainder of the paper we use as notation
\begin{align}\label{eq:alpha}
\alpha(\varphi):=\frac{1+\sin{(\varphi /2)}}{\sin{(\varphi /2)}}.
\end{align}

\begin{figure}[htb]
\begin{center}
  \includegraphics[width=.35\columnwidth]{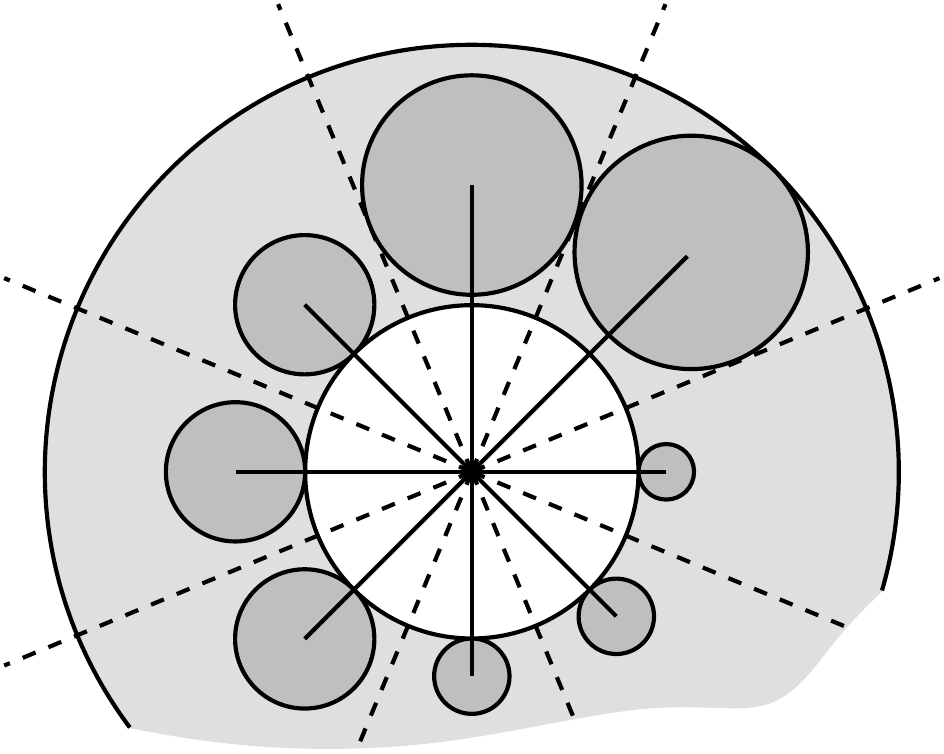} 
  \caption{A wedge layer (shaded) that had been filled with balloons by the greedy strategy.}
  \label{fig:layer}
\end{center}
\end{figure}

Notice that when a wedge situation occurs in round $i$, then in particular a wedge situation has to occur for the last balloon that is added in round $i$, since the balloons are sorted by increasing radii. All balloons placed in round $i$ are sandwiched between  $\text{SD}_i$ and $\text{SD}_{i+1}$. We call the region $\text{SD}_{i+1}\setminus \text{SD}_{i}$ the $i$-th layer $L_i$.\footnote{By convention $\text{SD}_1=\emptyset$, and for $i$ being the last round,  $\text{SD}_{i+1}=\text{smallest disk covering all balloons}$.} The \emph{width} of layer $L_i$ is defined as  $\safe_{i+1} - \safe_{i}$. When a wedge situation occurs in round $i$, the layer $L_i$ is called  a \emph{wedge layer}, otherwise a \emph{contact layer}. An example of a wedge layer is shown in Figure~\ref{fig:layer}. 



\subsection{Splitting the set of spokes} 
We come now back to the case where $n$ is not necessarily a power of two. In this setting there might be an odd number of spokes $k$ in some round. 
In such a round we place only  $\lfloor k/2 \rfloor$ balloons, such that no two of them are assigned to consecutive spokes.
This however has two drawbacks: First, the angles might not split evenly, and second, the layers will be filled with less balloons.

We can always pick $\lfloor k/2 \rfloor$ spokes such that in the remaining set of spokes at most two separating angles are smaller than the others, which are all equal.
Moreover, the two smaller angles are each at least half as big as the remaining angles. We call every set of spokes for which this property holds \emph{well-separated}. Furthermore we assume that  a well-separated set of spokes is ordered such that the two smaller angles are realized between the first and second, and between the second and third spoke. Algorithm~\ref{algo:spokes} describes a strategy that picks $\lfloor k/2 \rfloor$ of the spokes and ensures that the remaining set of spokes is still well-separated if the original set was well-separated.

\begin{algorithm}[H] 
\DontPrintSemicolon 
\SetKwInOut{Input}{Input}\SetKwInOut{Output}{Output} 
\Input{\small {$S$ set of spokes}}
\Output{\small $(T,T')$, such that $T'$ are the spokes that will be used in the current round,  $T=S\setminus T'.$}
\label{algo:spokes}
$T'$ $\leftarrow$  every spoke of $S$ with even index \;
$T\leftarrow S\setminus T'$ \label{line:spoke}\;
 reorder $T$ by putting the last spoke in front\;
 \Return  {$(T,T')$}\;
 \caption{ $\mathsf{SplitSpokes}(S)$}
\end{algorithm}

\begin{lem}\label{lem:spokes}
Let $S$ be a well-separated set of at least three spokes and let $\varphi$ denote the size of the big angles in $S$. Let $(T,T')$ be the return value of Algorithm~\ref{algo:spokes}.
\begin{enumerate}
\item[(1)] If $|T|>2$, then $T$ is well-separated.
\item[(2)] If $|T|=2$, then the smaller angle between the two spokes is at least $2\pi/3$.
\item[(3)] The wedge with angle $\varphi$ centered at the first spoke in $T'$ contains no spoke of $S$ in its interior.
\item[(4)] A wedge with angle $2\varphi$ centered at a spoke in $T'$ that is not the first spoke contains no spoke of $S$ in its interior.  
\end{enumerate}
\end{lem}
\proof
Let the angle between the first and second spoke in $S$ be $\gamma_1$, and let the angle between the second and third spoke in $S$ be $\gamma_2$. Since $S$ is well-separated, we have $\varphi/2\le\gamma_1,\gamma_2\le \varphi$. Hence the wedge centered at the second spoke of $S$ with angle $\varphi$ does not contain any other spoke of $S$ in its interior, which proves (3). Property (4) is due to the fact that every spoke in $S$ with even index larger than 2  is separated from its neighboring spokes by an angle of $\varphi$.
\begin{figure}[htb]
\begin{center}
\begin{tabular}{cp{.1cm}c}
  \includegraphics[width=.48\columnwidth]{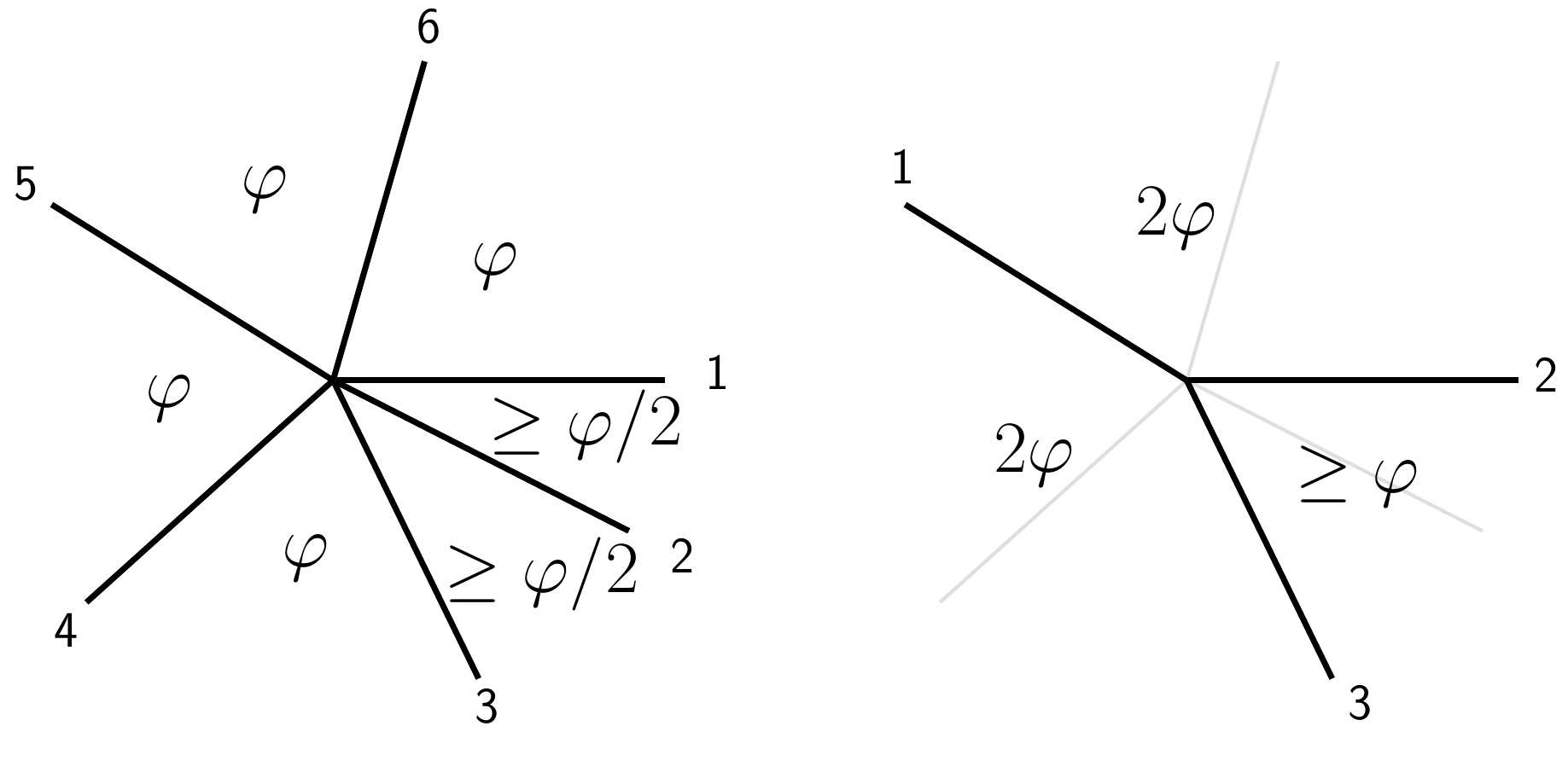} & &
  \includegraphics[width=.48\columnwidth]{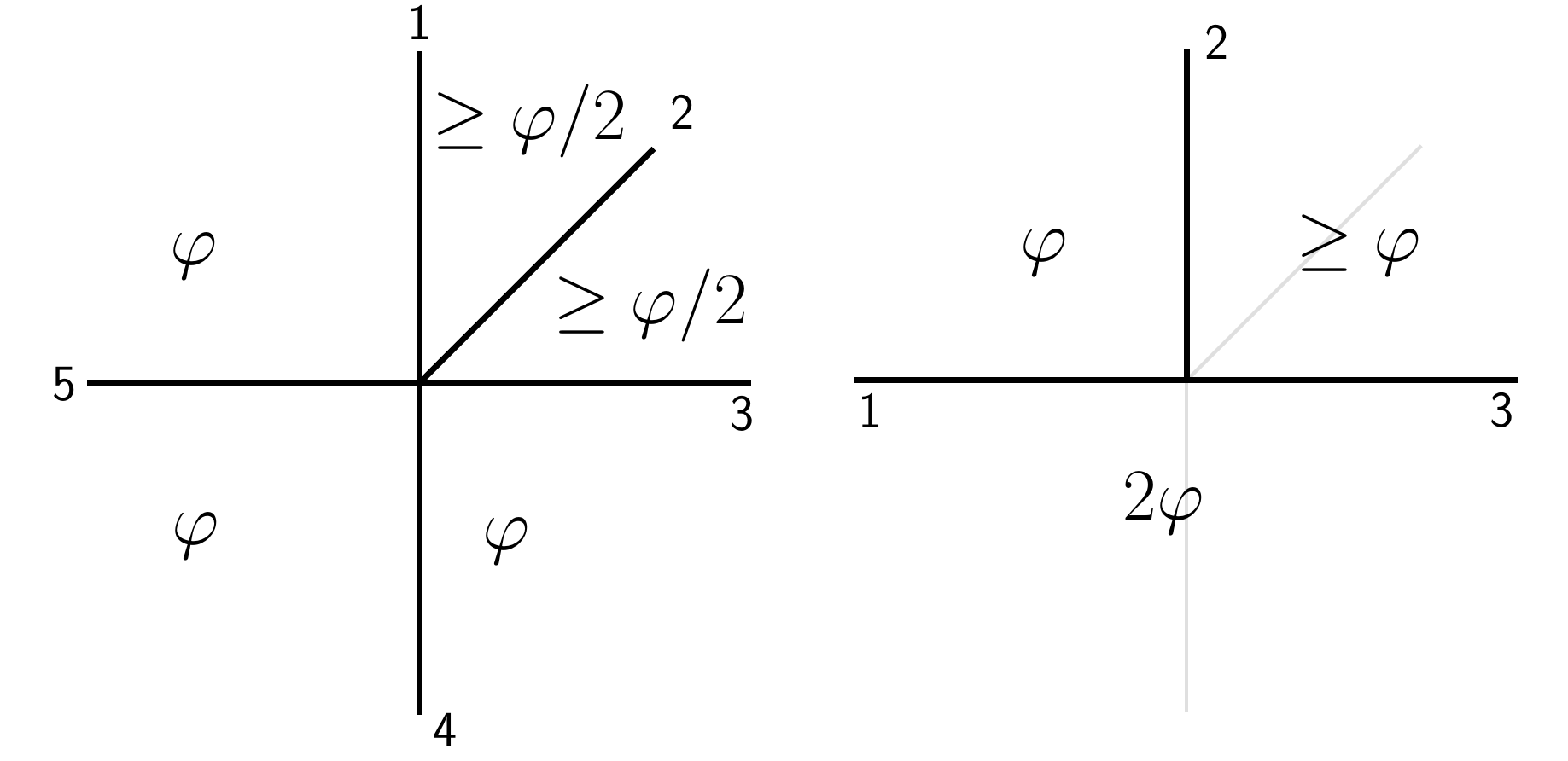} 
  \\
\small{(a)} & & \small{(b)} 
\end{tabular}
\caption{Merging the angles as implied by Algorithm~\ref{algo:spokes}. In case we have an even number of spokes (a), and in case we have an odd number of spokes (b). The spoke numbers are shown as small numbers.}
  \label{fig:spokes}
\end{center}
\end{figure}

After line~\ref{line:spoke} of Algorithm~\ref{algo:spokes}, the angle between the first and second spoke of $T$ equals $\gamma_1+\gamma_2\ge \varphi$. In case $S$ has an even number of spokes the remaining angles of size $\varphi$ are grouped pairwise and therefore the corresponding angles in $T$ are all $2\varphi$, which proves property (1) for this case. If the  set $S$ contains an odd number of spokes, the additional angle between the last spoke in $T$ and the first spoke in $T$ is also $\varphi$. Hence after reordering, the new set $T$ is again well-separated, and (1) follows. Figure~\ref{fig:spokes} illustrates the outcome of Algorithm~\ref{algo:spokes}. 

To see that (2) is true, notice the following. $T$ contains two spokes, if $S$ contains three or four spokes. In case $S$ contains $4$ spokes, the sum of the two small angles is at least $2\pi/3$. In case $S$ contains three spokes, the sum of the two small angles between the spokes is  at least $\pi$. The large angle between the spokes in $S$ is at least $2\pi/3$. This angle appears also between the spokes in $T$.
\endproof

To ensure that the balloons of each layer cannot interfere with each other and with the remaining spokes, we place them inside the wedges defined by Lemma~\ref{lem:spokes}(3--4). All wedges have the same opening angle, say $\varphi$, except the first wedge, whose opening angle is at least $\varphi/2$. The balloon with the smallest radius in each round is placed inside the wedge with the (possible) smaller opening angle.

\subsection{The final layer}

It is important to analyze the situation where the greedy strategy has to stop. In every round we reduce the number of spokes from $k$ to $\lceil k/2 \rceil$. If we subdivide the spokes in this fashion we will come to a point where exactly two spokes are left. The final two balloons are placed in the last round as follows: (1) The balloon $B_n$ will be placed such that it touches the safe disk. (2) The balloon $B_{n-1}$ will be placed such that it is contained inside a wedge with opening angle $\pi/3$, centered at its spoke, while avoiding the interior of the current safe disk.

\begin{lem}\label{lem:basecase}
When the balloons are placed as discussed in the previous paragraph, then one of the following is true:
\begin{enumerate}
\item The width of the last layer is $2 r_n$.
\item All balloons can be covered with a disk of radius two.
\end{enumerate}
\end{lem}
\proof
Let $\varphi$ be the smaller of the two angles between the spokes in the final round $i$. 
Due to Lemma~\ref{lem:spokes}, $\varphi$ is at least $2\pi/3$. The tangent of $B_n$ at its intersection with $\text{SD}_i$ separates $B_n$ from the spoke of $B_{n-1}$. Since the angle between this tangent and the spoke of $B_{n-1}$ is at least $\varphi-\pi/2\ge \pi /6$ it is safe to place $B_{n-1}$ inside a wedge centered at its spoke with opening angle $\pi/3$. Thus, either $B_{n-1}$ touches $\text{SD}_i$, or it is contained inside a disk of radius $\alpha(\pi/3)r_{n-1}=3r_{n-1}$. In the former case the width of the layer is  $2 r_n$, in latter case the radius of the covering disk is at most $\max\{2 r_n, 3/2\}$ (recall that $r_{n-1}\leq 1/2$).
\endproof

%
Due to Lemma~\ref{lem:basecase} we can assume that the width of the last layer equals $2 r_n$. Thus even if $B_{n-1}$ defines a wedge situation we  consider the last layer as contact layer.
We summarize the discussion in Algorithm~\ref{algo:greedy2}.

\begin{algorithm}  
\caption{$\mathsf{GreedyBalloon(S)}$.}
\DontPrintSemicolon 
\SetKwInOut{Input}{Input}\SetKwInOut{Output}{Output} 
\label{algo:greedy2}
\Input{\small {$S\colon \text {spokes in cyclic order.}$   }}
 $k\leftarrow 0$ \tcp*{number of balloons placed so far}  
 $\safe\leftarrow 0$ \tcp*{radius of the current safe disk} 
\While{$|S|>2$}{
$(T,T')\leftarrow \mathsf{Splitspokes}(S)$ \;
$\text{width}\leftarrow 0$  \tcp*{width of the current layer so far} 
\For{$i\leftarrow k+1$ {\bf to} $k+|T'|$}{
$s\leftarrow$ $(i-k)$-th spoke of $T'$\;
$\varphi\leftarrow 2 (\text{minimal angle between $s$ and one of its neighboring spokes in $S$})$\;
 $c\leftarrow \max\left\{ \alpha(\varphi) r_i-r_i  , \safe +  r_i \right\}$ \tcp*{center of $B_i$}
place $B_i$ on $s$ at distance $c$ \;
$\text{width}\leftarrow \max\{\text{width}, c+r_i-\safe\}$\;
}
 $\safe\leftarrow\safe+\text{width}$\;
$k \leftarrow k + |T'|$ \;
$S \leftarrow T$ \;
}
let $s_1,s_2$ be the spokes in $S$ \;
 place $B_n$ on  $s_1$ at distance $\safe+r_n$\;
 place $B_{n-1}$ on  $s_2$ at distance $\max\{2 r_{n-1}, \safe+r_{n-1}\}$\;

%
\end{algorithm}


\subsection{Quality of the greedy strategy}

We denote by $R$ the  radius of the smallest disk  that covers all balloons. 
In order to determine $R$ we have to consider only certain radii.
\begin{lem}\label{lem:1}
The radius of the smallest disk $R$ that covers all balloons drawn with Algorithm~\ref{algo:greedy2} can be determined with the knowledge of
\begin{enumerate}
\item the number of spokes,
\item the radius of the largest and smallest balloon in the outermost wedge layer, 
\item the radii of the largest balloons in each of the contact layers following the outermost wedge layer.
\end{enumerate} 
\end{lem}
\proof
Suppose the last wedge situation occurs in round $i$. Then the radius  of $\text{SD}_{i+1}$ is determined by a balloon that touches its wedge.
All wedges have the same opening angle, except maybe the first wedge. Since the smallest balloon is placed inside the first wedge, the wedge situation that defines the radius  of $\text{SD}_{i+1}$ depends on the possible wedge situation of the largest and smallest balloon only.
The following layers are all contact layers. Their width is determined by the diameter of the largest balloons in each layer. The radius $R$ equals  therefore the radius of $\text{SD}_{i+1}$ with the addition of the widths of the following contact layers.
\endproof

Since we are interested in a worst case bound for $R$ we  make the following assumptions to simplify the analysis of the algorithm.
\begin{lem}\label{lem:2}
Let $r_w$ be the radius of the balloon, whose wedge situation determined the width of the last wedge layer $L_k$. The radius $R$ of the smallest covering disk is maximized when
\begin{align*}
r_{w}&=r_{w+1}=r_{w+2}=\cdots=r_{n-1}, \text {and} \\
r_{1}&=r_{2}=r_{3}=\cdots=r_{w-1}=0.
\end{align*}
\end{lem}
\proof
We consider the radii as  resources that we want to spend to make $R$ as large as possible. Since no radius of a balloon with smaller index than $w$ matters for $R$, we  set these radii to zero to save resources. If $B_w$ is the smallest balloon in its layer, all radii of balloons in $L_k$ have the same radius in the worst case. Otherwise we could shrink some of these balloons without changing the width of $L_k$ and spent the resources to increase $r_n$ and therefore $R$.

Only the balloon added last in each contact layer determines the width of its layer. We select the radii of the other balloons in contact layers as small as possible, i.e., as large as the radius of the largest balloon in the previous layer.  If any of these radii would be larger we could make such a radius smaller and increase $r_n$ instead, which would increase $R$. 

Assume we have at least two contact layers following $L_k$.
Let $B_c$ be the largest  balloon in the contact layer $L_{k+1}$, that is the balloon last added in $L_{k+1}$. Due to the discussion in the previous paragraph we can assume that the balloon $B_{c+1}$ in the next layer has  radius $r_c$.
If $r_c>r_w$, we could lower the radius  by $r_c-r_w$ for $B_{c}$ and $B_{c+1}$ each. By this we can increase $r_n$ by $2(r_c-r_w)$. 
As a consequence the radius $R$ increases by $r_c-r_w$. Therefore in the worst case all radii in layer $L_{k+1}$ equal $r_w$.
By an inductive argument   the radii in the last contact layers are all $r_w$.
The only exception is the largest balloon $B_n$.
\endproof


\begin{thm}\label{thm:general}
Algorithm~\ref{algo:greedy2} constructs a drawing of balloons with disjoint interiors and spokes that intersect only the interior of their associated balloon that can be covered with a disk of radius two, which is best possible in the worst case. 
\end{thm} 
\proof
We define as $\bar L_i$  the $i$-th last layer such that $\bar L_1$ is the last layer.
Suppose there were $\ell$ spokes left, before the last wedge layer was filled. We denote the number of contact layers that follow the last wedge layer by $k$. By Algorithm~\ref{algo:spokes} the number $k$ is given by a function $k= f(\ell)$, which is  defined as follows
\begin{align}\label{equ:k}
f(\ell):= \begin{cases}
1 & \text{if $3\le \ell\le4$,} \\
1+f\left(\frac{\ell}{2}\right) & \text{if $\ell>4$, even,}  \\
1+f\left(\frac{\ell+1}{2}\right) & \text{if $\ell >4$, odd.} 
\end{cases}
\end{align}
By induction, $f(\ell) \le \log (\ell-1)$. The radius of the covering disk $R$ equals the radius of $\bar L_{k}$'s safe disk plus the width of the last $k$ contact layers. Let $B_w$ be the balloon that determined $\safe_{k}$. By Lemma~\ref{lem:2} we can assume that all balloons following $B_w$ have radius $r_w$, except $B_n$. All other radii are zero.

As previously discussed, the balloon $B_w$ is either the first or the last balloon in the last wedge layer. We discuss the two possibilities by case distinction. Let us first assume that  $B_w$ is the last balloon of layer $\bar L_{k+1}$. By construction the last balloon is placed inside the wedge with largest opening angle (in this round). Therefore its  opening angle $\varphi$ is minimized, when the angles between all pairs of neighboring spokes are equal.
We have $\ell$ spokes in $\bar L_{k+1}$, and therefore two spokes are separated by $2\pi/\ell$ and $\varphi=4\pi/\ell$. Furthermore, we have $k-1$ layers of width $2 r_w$, and one layer of width $2r_n$ following $\bar L_{k+1}$.
In layer $\bar L_{k+1}$ we place no more than $ \ell/2$ balloons and therefore in the last $k$ layers we have at least $\ell/2$ balloons in total. Since there is one balloon in $\bar L_{k+1}$ with radius $r_w$ and only one balloon in the last $k$ layers with radius different from $r_w$, we get $r_n\le1- r_w \ell/2$.
This leads to 
\begin{align*}
 R& \le \alpha(\varphi) r_w+2(k-1)r_w+2r_n \le 2+\left[ \alpha(4\pi/\ell) +2\log(\ell-1)-\ell-2 \right] r_w.
 \end{align*}
The last wedge layer must contain  at least three spokes. Since $\alpha(4\pi/\ell) +2\log(\ell-1)-\ell-2$ is decreasing\footnote{The estimation of this expression and of similar following expressions was obtained by computer algebra software.}  for $\ell\ge 4$ and negative for $\ell=3,4$, we get $R\leq2$.

We assume now that $B_w$ was placed first in $\bar L_{k+1}$. Again, let $\varphi$ be the angle of the wedge that contains $B_w$ centered at its spoke. Due to Lemma~\ref{lem:spokes} the angles between two neighboring spokes are all of size $\psi$ except two angles, which are at least $\psi/2$ (the small angles). The angle $\varphi$ is twice the minimum of the two small angles, and hence minimized when one of the small angles has size $\psi$ and the other has size $\psi/2$. In this case we have $\ell-1$ angles of size $\psi$ and one angle of size $\psi/2$. Since all angles sum up to $2\pi$, we have $\psi=2\pi/(\ell-1/2)$, which is a lower bound for $\varphi$. Notice that all balloons in $\bar L_{k+1}$ have now radius $r_w$, hence we have $\ell-1$ balloons of radius $r_w$, and therefore $r_n\le1-(\ell-1)r_w$. We conclude with
 \begin{align*}
 R& \le \alpha(\varphi) r_w+2(k-1)r_w+2r_n\le 2+\left[ \alpha(2\pi/(\ell-1/2)) +2\log(\ell-1)-2\ell  \right] r_w.
 \end{align*}
 For $\ell\ge2$ the expression  $\alpha(2\pi/(\ell-1/2)) +2\log(\ell-1)-2\ell $ is negative and decreasing
  and the theorem follows.
\endproof

%
%
%
%
%
%
%

\section{Free spokes}

In this section we discuss a variant of the balloon layout problem that finds application in a tree drawing algorithm, which is presented in Section~\ref{sec:drawing}. In contrast to the original setting we require that one or two spokes remain without balloon. Hence the number of spokes exceeds the number of balloons which we denote with $n$. A spoke that remains without balloon is called \emph{free spoke}. As additional constraint we require that if there are two free spokes, the smaller separating angle is at least $2\pi/3$. Allowing free spokes makes the performance of the greedy strategy worse, since the available angular space  between the spokes is reduced. 
In order to achieve  good bounds for this modified problem, we change the greedy strategy slightly. In particular, we change the terminal cases for the layout algorithm and we introduce a 
construction that allows us to move some balloons inside their safe disk. The rest of the greedy strategy remains unaltered.

\subsubsection*{New terminal cases}
We have two terminal cases for the scenario with one free spoke, and two terminal cases for the scenario with two free spokes. The new terminal cases are covered by the Lemmata~\ref{lem:base-3s2b}---\ref{lem:base-4s2b}; see also Table~\ref{tab:terminalcases}. Notice that for every number of original spokes the greedy strategy has to come to one of these terminal cases. 

\renewcommand\arraystretch{1.2}%
\begin{table}[htbp]
  \centering
  \begin{tabular}{@{} ccc @{}}
   \hline \hline
    ¥ & remaining spokes & remaining balloons\\ 
\hline\hline
    \multirow{2}{*}{Lemma~\ref{lem:base-3s2b}} & 2 & 1 \\ 
     & 3 & 2 \\  \hline
    Lemma~\ref{lem:base-3s1b}  & 3 & 1 \\ \hline
    Lemma~\ref{lem:base-4s2b}  & 4 & 2 \\ 
 \hline
      \end{tabular}
  \caption{The terminal cases.}
  \label{tab:terminalcases}
\end{table}

%

\begin{lem}\label{lem:base-3s2b}
Suppose we have either two spokes and one balloon, or three well-separated spokes and two balloons left while executing the greedy strategy. We can place the remaining balloons, such that either all balloons can be covered with a disk of radius two, or the width of the last layer is $2r_n$. 
\end{lem}
\proof
The case when there are two spokes and one balloon left is trivial. For the remaining case we assume that 
 the spokes are labeled such that  the largest angle is realized between $s_3$ and $s_1$, and the second largest angle is realized between $s_2$ and $s_3$. Balloon $B_n$ is placed at  $s_3$ such that it touches the safe disk, which is possible, since the  angle between  $s_2$ and $s_3$ is at least $\pi/2$. Let $t$ be the tangent of $B_n$ at the intersection with the safe disk. Since $s_3$ and $s_1$ are separated by an angle of at least $2\pi/3$, $t$ and $s_1$ are separated by an angle of at least $\pi/6$. Thus it is safe to place $B_{n-1}$ at a wedge with opening angle $\pi/3$ centered at $s_1$. If this would result in a wedge situation, the disk covering all balloons except possibly $B_n$ would have radius $\alpha(\pi/3)r_{n-1}=3r_{n-1}<2$. 
\endproof

\begin{lem}\label{lem:base-3s1b}
Suppose we have three well-separated spokes and one balloon left while executing the greedy strategy. We can place $B_n$, such that  the width of the last layer is $(1+\sqrt{2-2/\sqrt{5}} ) r_n$,  and the smaller angle between the two remaining spokes is at least $2\pi/3$. 
\end{lem}
%
%
\proof
By Lemma~\ref{lem:spokes} the smallest angle between two spokes is at least $2\pi/5$. Thus we can place $B_n$ on the spoke incident to the two smaller angles such that it touches the safe disk inside a wedge with opening angle $4\pi/5$. 
Hence, $B_n$ can be covered with a disk of radius $\alpha(4\pi/5)r_n=(1+\sqrt{2-2/\sqrt{5}}) r_n\approx 2.0514r_n$.
The remaining spokes are separated by the former larger angle. Again, by Lemma~\ref{lem:spokes} this angle is at least $2\pi/3$ and at most $\pi$.
\endproof

\begin{lem}\label{lem:base-4s2b}
Suppose we have four well-separated spokes and two balloons left while executing  the greedy strategy. We can place $B_n$ and $B_{n-1}$, such that either all balloons can be covered with a disk of radius two, or the width of the last layer is $2r_n$. The smaller angle between the two remaining spokes is at least $2\pi/3$. 
\end{lem}
\proof
By well-separatedness, we can assume that the two larger angles (which are at least $\pi/2$) are realized between the spokes $s_3$, $s_4$, and $s_1$. We place $B_n$ at $s_4$ such that it touches the safe disk. The smallest angle is minimized when all other angles are equal. In this case, the smallest angle is $2\pi /7$.
Hence, we can place $B_{n-1}$ at $s_2$ inside a wedge with opening angle $4\pi/ 7$. If this would result in a wedge situation, the disk covering all balloons would have radius $\alpha(4 \pi/7)r_{n-1}\le1.139..<2$. The angle between the two remaining (free) spokes is at least the sum of the two small angles, which is at least $2\pi/3$.
\endproof

\subsubsection*{Compactification}

The following construction allows us to place a balloon such that it slightly overlaps the
previous safe disk; this is needed in a few special cases.

\begin{figure}[htb]
\begin{center}
  \includegraphics[width=.32\columnwidth]{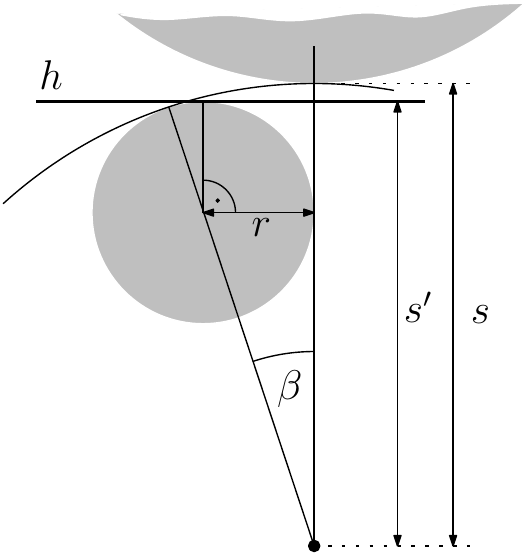} 
 \caption{By Lemma~\ref{lem:overlap} it is possible to push the balloons slightly inside the safe disk.}
 \label{fig:push}
\end{center}
\end{figure}

\begin{lem}\label{lem:overlap}
Suppose that $s_1$ and $s_2$ are two spokes that are separated by an angle $\beta$ and
that $B$ is a balloon placed on $s_1$ such that it is disjoint from $s_2$ and
such that it can be covered with a disk of radius $s$.
Then a balloon $B'$ placed on $s_2$ at distance 
$s\cdot(\sin(\beta) + \cos(\beta))/(\sin(\beta) + 1)$ from the origin will be disjoint from $B$.
\end{lem}
\proof
In the worst case $B$ is as large as possible, i.e., it touches both,
the spoke $s_2$ and the border of the disk of radius $s$. Let the radius of $B$
in this case be $r$. By Lemma~\ref{lem:2} we have $s = r + r/\sin(\beta)$ (see Figure~\ref{fig:push})
and hence $r = s\cdot\sin(\beta)/(\sin(\beta) + 1)$.
To ensure that $B'$ is disjoint from $B$, it suffices to place it above the line $h$
that is perpendicular to $s_2$ and touches $B'$. The distance of this line
from the origin is 
\[
r + \cot(\beta)r = r\cdot (\sin (\beta) + \cos(\beta))/\sin (\beta)
= s \cdot (\sin (\beta) + \cos(\beta))/( \sin(\beta) + 1)
.
\]
\endproof

\subsubsection*{Analysis of the modified greedy strategy}
The analysis follows the presentation in Section~\ref{sec:basic}. 
As before, the  layer of the last round is always considered as contact layer, even when a wedge situation determined its width. 

\begin{thm}\label{thm:1left}
Assume that the number of spokes exceeds the number of balloons by one.
Algorithm~\ref{algo:greedy2} with  base case  as described in Lemma~\ref{lem:base-3s2b} produces a drawing of balloons with disjoint interiors and one free spoke that can be covered with a disk of radius two. 
\end{thm}
\proof
We denote by $\ell$ to be the number of spokes in the last wedge layer. We reuse the estimations for the angles between the spokes provided in the proof of Theorem~\ref{thm:general}. There are $k$ layers following the last wedge layer. The number $k=f(\ell)$ can bounded in terms of $\ell$ by the following recursion
\begin{align*}
f(\ell):= \begin{cases}
1 & \text{if $4\le \ell \le 6$,} \\
1+f\left(\frac{\ell}{2}\right) & \text{if $\ell>6$, even,}  \\
1+f\left(\frac{\ell+1}{2}\right) & \text{if $\ell >6$, odd.} 
\end{cases}
\end{align*}
The recursion yields $f(\ell)\le\log(2(\ell-1)/3)$, which can by checked by induction. 

Let $B_w$ the balloon that determined the width of the last wedge layer. $B_w$ can be either the first or last balloon of the layer. Assume that $B_w$ was placed last in $\bar L_{k+1}$. In this case $r_n\le1-(\ell/2-1)r_w$, since we have one balloon less compared to the proof of Theorem~\ref{thm:general}, but we have the same bounds for the angles, namely $\varphi\ge 4\pi / \ell$. This gives
\begin{align*}
 R& \le \alpha(\varphi) r_w+2(k-1)r_w+2r_n \le 2+\left[ \alpha(4\pi/\ell) +2\log(2(\ell-1)/3)-\ell \right] r_w.
\end{align*}
Since $\alpha(4\pi/\ell) +2\log(2(\ell-1)/3)-\ell$ is non-positive for  all $\ell\ge 4$, we have $R\le 2$ in this case.

Assume now that $B_w$ was the first balloon of $\bar L_{k+1}$. By the proof of Theorem~\ref{thm:general} we have   $\varphi\ge 2\pi/ (\ell-1/2)$. Since we have one balloon less, we get $r_n \le 1-(\ell-2)r_w$. We deduce 
\begin{align*}
 R& \le \alpha(\varphi) r_w+2(k-1)r_w+2r_n  \\
 &\le 2+\left[ \alpha(2\pi/(\ell-1/2)) +2\log(2/3(\ell-1))-2\ell+2  \right] r_w.
\end{align*}
Since $ \alpha(2\pi/(\ell-1/2)) +2\log(2/3(\ell-1))-2\ell+2<0$ for $\ell \ge 4$ the theorem follows.
\endproof

\begin{thm}\label{thm:2left}
Assume that the number of spokes exceeds the number of balloons by two.
Algorithm~\ref{algo:greedy2} with base cases as described in Lemma~\ref{lem:base-3s1b}, and~\ref{lem:base-4s2b}, and the construction described in Lemma~\ref{lem:overlap} produces a drawing of balloons with disjoint interiors and two free spokes that can be covered with a disk of radius $(1+\sqrt{2-2/\sqrt{5}})=2.0514..$. 
\end{thm}
\proof
The proof is similar to the proof of Theorem~\ref{thm:1left}. So again, let $\ell$  be the number of spokes in the last wedge layer. 
We stop the greedy strategy when three or four spokes are left. 

%
%
Let $k$ denote the numbers of layers following the last wedge layer.
We have \begin{align*}
k= f(\ell):= \begin{cases}
1 & \text{if $5 \le \ell\le8$,} \\
1+f\left(\frac{\ell}{2}\right) & \text{if $\ell>8$ even,}  \\
1+f\left(\frac{\ell+1}{2}\right) & \text{if $\ell >8$ odd.} 
\end{cases}
\end{align*}
The solution to this recurrence gives $f(\ell)\le \log(\ell-1) -1 $, which can be checked easily by induction.

Assume that $B_w$ is the last balloon in its layer. We place $\lfloor \ell/2 \rfloor$ balloons in the last wedge layer, and therefore $\lceil \ell/2  \rceil - 2$ in the final $k$ layers. This gives $r_n\le 1 - (\lceil \ell/2  \rceil - 2)r_w$. Let $\bound:=\left(1+\sqrt{2-2/\sqrt{5}}\right)$, by Lemma~\ref{lem:base-3s1b} the width of the last layer is at most $\bound\cdot  r_n$. We obtain
\begin{align*}
 R& \le \alpha(\varphi) r_w+2(k-1)r_w+ \bound r_n \\
 &\le \bound+\left[ \alpha(4\pi/\ell) +2\log(\ell-1)-4+\bound (2-\lceil \ell/2  \rceil) \right] r_w.
\end{align*}
A numerical analysis shows that $R<\bound$ when $\ell\ge7$. Thus in the two remaining cases ($\ell=5,6$) we apply
Lemma~\ref{lem:overlap} to enhance the result by moving $B_n$ slightly inwards. In both cases, the last layer $\bar L_{1}$
contains only $B_n$, the last wedge layer is $\bar L_{2}$, and we have $r_n \le 1 - r_w$.
The angle between two spokes in layer $\bar L_{2}$ is at least $2\pi/(2\ell-1)$, so we
can use $\beta = 2\pi/(2\ell-1)$ in Lemma~\ref{lem:overlap}. In this way, we obtain
\begin{align*}
R &\le \alpha(\varphi)r_w\cdot (\sin (\beta) + \cos(\beta))/( \sin(\beta) + 1) + \bound r_n\\
&\le  \bound + [\alpha(4\pi/\ell)\cdot (\sin (\beta) + \cos(\beta))/( \sin(\beta) + 1) - \bound]r_w.
\end{align*}
This is less than $\bound$ in both cases, $\ell = 5$ and $\ell = 6$.

Finally we have to consider the case when  $B_w$ is the smallest balloon in its layer. In this setting we have $r_n\le 1-(\ell-3)r_w$ and $\varphi\ge 2\pi/(\ell-1/2)$, which yields
\begin{align*}
 R& \le \alpha(\varphi) r_w+2(k-1)r_w+\bound\cdot r_n \\
 &\le \bound+\left[ \alpha(2\pi/(\ell-1/2)) +2\log(\ell-1)-4-\bound( \ell-3)  \right] r_w.
\end{align*}
We obtain,  $R\le \bound$, since $\alpha(2\pi/(\ell-1/2)) +2\log(\ell-1)-4-\bound( \ell-3)<0 $  for $\ell \ge 5$.
 \endproof

\section{Drawing unordered trees with perfect angles}\label{sec:drawing}


The greedy strategy can be used to construct drawings of unordered trees with perfect angular resolution and small area. In fact, the balloon layout problem studied in Section~\ref{sec:basic} is a subproblem of the drawing algorithm of Duncan~\etal~\cite{DEGKN10}, where it is used to draw depth-1 trees. With the help of the so called heavy edge tree-decomposition (see Tarjan~\cite{T83}) these trees are combined to the original tree. Since our proposed strategy uses significantly smaller area, it  implies  an improvement for the area of the tree drawing.

\begin{figure}[htb]
\begin{center}
  \includegraphics[width=.6\columnwidth]{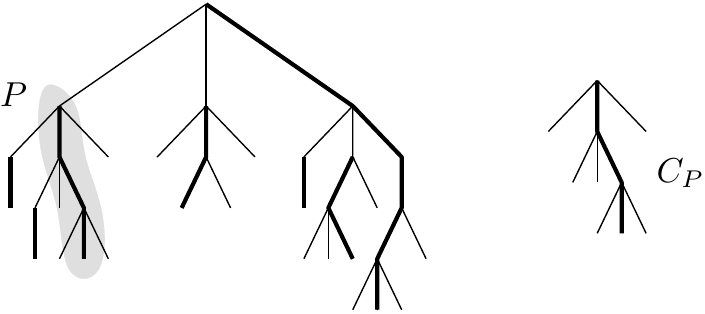} 
 \caption{An example of a heavy-edge tree-decomposition. The path $P$ has depth two.}
  \label{fig:heavy}
\end{center}
\end{figure}

We start with a brief review of the  heavy edge tree-decomposition. 
Let $u$ be a non-leaf of the rooted tree $T$ with root $r$. We denote by $T_u$ the subtree of $T$ rooted at $u$. Let $v$ be the child of $u$ such that $T_v$ has the largest number of nodes (compared to the subtrees of the other children of $u$), breaking ties arbitrarily. We call the edge $(u,v)$ a \emph{heavy edge}, and the edges to the other children of $u$ \emph{light edges}. The heavy edges induce a decomposition of $T$ into (maximal) paths, called \emph{heavy paths}, and light edges; see Figure~\ref{fig:heavy} on the left. 
We call the  node on a heavy path that is closest to $r$ its \emph{top node}. The subtree induced by a heavy path is the subtree rooted at its top node. The light edge that links the top node with its parent in $T$ is called \emph{light parent edge}. The \emph{depth} of a heavy path $P$ is defined as follows: If $P$ is not incident to light parent edges of other heavy paths it has depth one. Otherwise we obtain the  depth of  $P$ by adding one to the maximal depth of a heavy path linked to $P$ by its light parent edge. 

%

%
\begin{figure}
\centering
  \includegraphics[width=.65\columnwidth]{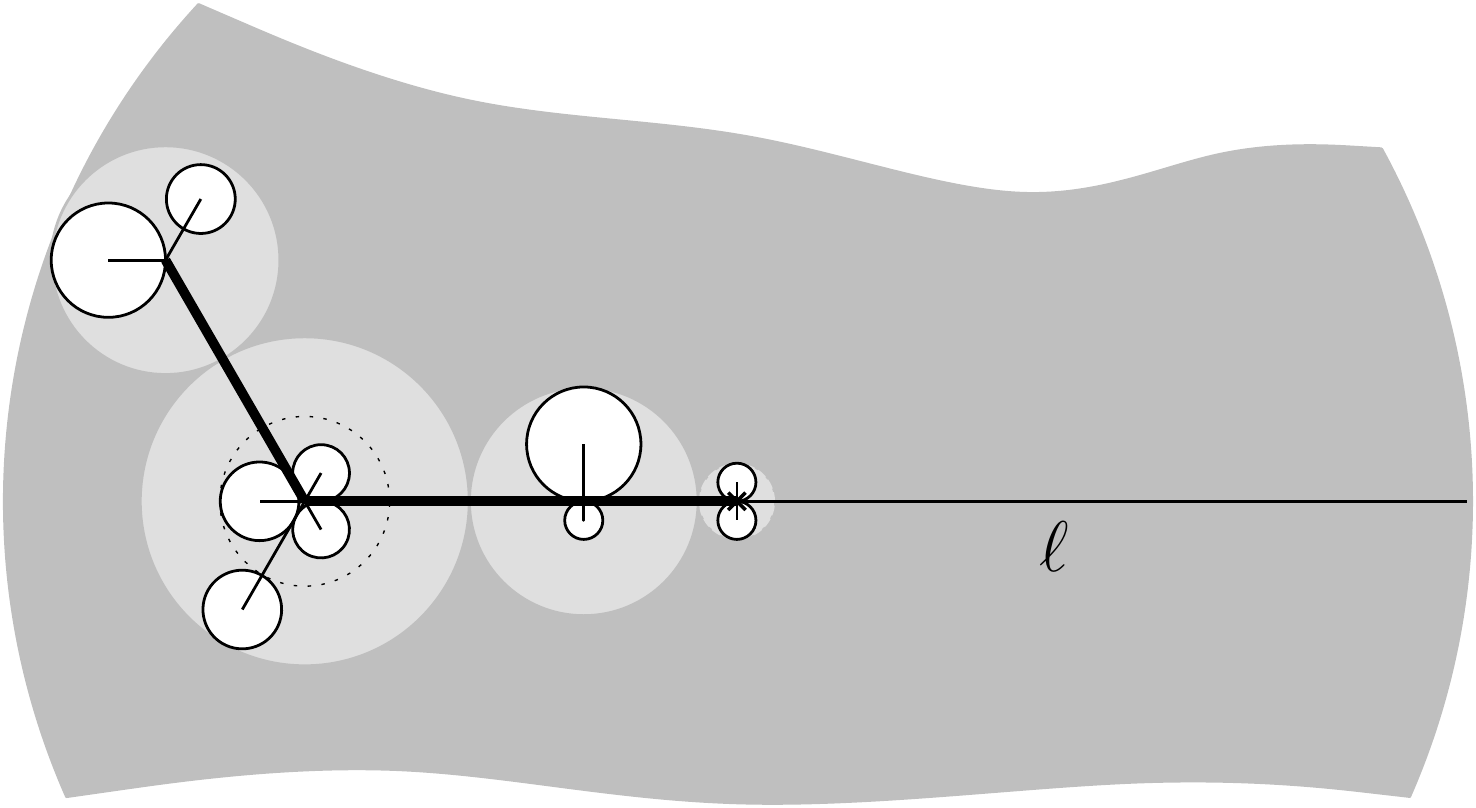} 
 \caption{Constructing the drawing for the subtree of a heavy path with light parent edge $\ell$. Its exclusive disk is drawn darkly shaded. The covering disks of the balloon layouts around the heavy path nodes are lighter shaded. White disks indicate the exclusive disks of heavy tree subtrees with smaller depth.}
  \label{fig:rootpath}
\end{figure}

We apply recursion to draw the tree. In particular, we construct drawings of the subtrees of all heavy paths including their light parent edge. For every such subtree $S$ we define an \emph{exclusive disk} $X$ with the following properties:  (1) The drawing of $S$ is contained inside $X$, (2) the light parent edge of $S$ crosses the boundary of $X$ orthogonally, and (3) the center of $X$ coincides with the top node of the corresponding heavy path.  

Let $P$ be a heavy path of  $T$ and let $C_P$ be the union of $P$  with its incident light edges (but without the light parent edge) as shown in Figure~\ref{fig:heavy}. The leaves of $C_P$ represent (possibly degenerate) subtrees of heavy paths with smaller depth. 
Assume we have constructed the drawings for all these subtrees and we constructed an exclusive disk for each of the drawings. Let $u$ be a non-leaf tree node of $P$. We apply the balloon layout algorithm to draw a tree rooted at $u$. We introduce a spoke for every light edge incident to $u$ and the balloon that is placed on a spoke represents the exclusive disk of the corresponding subtree. The heavy edges incident to $u$ are represented as free spokes.  In order to combine the balloon layouts for the nodes on $P$ we apply the construction of Duncan~\etal\cite[full version, Lemma 2.3]{DEGKN10}. This Lemma states that the combined drawing fits inside an exclusive disk of radius $2\sum_i x_i$, where $x_i$ is the radius of the disk that covers the balloon layout of the $i$-th node on $P$. Figure~\ref{fig:rootpath} illustrates this construction. 
The terminal step in the recursion draws the leaves of $T$ (degenerate heavy paths) with their incident light parent edges. This is done by drawing these light parent edges with length one and place the exclusive disk centered at the leaf node with radius one.
The following Theorem presents a bound on the area of the constructed tree drawing.

\begin{thm}\label{thm:main}
Let $\bound=(1+\sqrt{2-2/\sqrt{5}})$ be the constant derived in Lemma~\ref{lem:base-4s2b}. Using Algorithm~\ref{algo:greedy2} in the framework of Duncan~\etal~produces a drawing of an unordered tree with $n$ nodes that has perfect angular resolution and that can be covered with a disk of radius $ n^2\cdot n^{\log \bound}< n^{3.0367}$, while having no edge with length smaller than 1.
\end{thm}
\proof
Let $N_u$ denote the number of nodes in the subtree rooted at $u$.
We show  by induction that the area of a subtree of a heavy path $P$ of depth $i$ is less than $(2\bound)^iN_u$, for $u$ being the top node of $P$. This statement is certainly true for $i=1$. 
Assume that we have already built the drawings of all depth $(i-1)$  heavy path  subtrees. We apply the construction of the previous paragraph to combine the drawings. In order to achieve this we have to apply the greedy strategy for the balloon layout for every node of $P$. By Theorem~\ref{thm:1left} and~\ref{thm:2left}, the balloon layout requires a covering disk of radius $\bound$ times the sum of the balloon radii. We denote the necessary radius of the covering disk at note $z$ by $x_z$, and the number of nodes of the subtrees that are linked to $z$ by a light parent edge by $M_z$. By the recursion hypothesis we have $x_z\le \bound (2\bound)^{(i-1)}M_z$.  The construction of Duncan~\etal\cite[full version, Lemma 2.3]{DEGKN10} combines the balloon layouts of the nodes of $P$ to a tree drawing with perfect angular resolution and the drawing fits inside an exclusive disk of radius at most $2\sum_{z \in P}x_z=\sum_{z \in P} (2\bound)^iM_z=(2\bound)^iN_u$.

Since every root-leaf path in $T$ traverses at most $\log n$ light edges, the  depth of the root of $T$ is at most $\log n$. This shows that the area of the complete drawing is at most \[(2\bound)^{\log n}N_r=n^{\log 2\bound} \cdot n=n^2 n^{\log \bound}< n^{3.0367}.\]

Notice that by construction all edges have length at least one.
\endproof

\section{Concluding remarks}

The algorithm presented in this paper runs in linear time. Even when the set of balloons $\mathcal{B}$ is not ordered by radii we can obtain a running time of $O(n)$ for Algorithm~\ref{algo:greedy2}. The correctness of  the algorithm follows from a weaker condition. Recall that $r_i$ denotes the radius of the balloon $B_i$. We denote the median of the radii of $\mathcal{B}$ with $\bar r$.
\begin{defn}[weakly ordered sequence ]
We say that the sequence of balloons $\mathcal{B}$ is \emph{weakly-ordered}, iff
\begin{enumerate}
\item $r_1=\min\{r_i\}$,
\item for all $i< \lfloor n/2\rfloor$ we have $r_i\le \bar r$,
\item $r_{\lfloor n/2\rfloor}=\bar r$, and 
\item $(B_{\lfloor n/2\rfloor+1},\dotsc,B_n)$ is weakly-ordered.
\end{enumerate}
\end{defn}
It follows from the recursive definition that $\mathcal{B}$ can be weakly-ordered in linear time. On the other hand it is indeed sufficient for $\mathcal{B}$ being weakly-ordered, since for the location of the balloons within each round only the smallest and largest balloon matters. A permutation of the  balloons in between has no influence of the necessary width of the corresponding layer.

The only case, where we obtain no strict inequalities in the proof of Theorem~\ref{thm:general}, is when $|\mathcal{B}|=1$. By placing all balloons slightly  inside the wedges, resp., slightly outside the safe disks we can therefore modify all constructions such that no balloons touch.

\begin{figure}[ht]
  \begin{center}
    \includegraphics[width=0.29\textwidth]{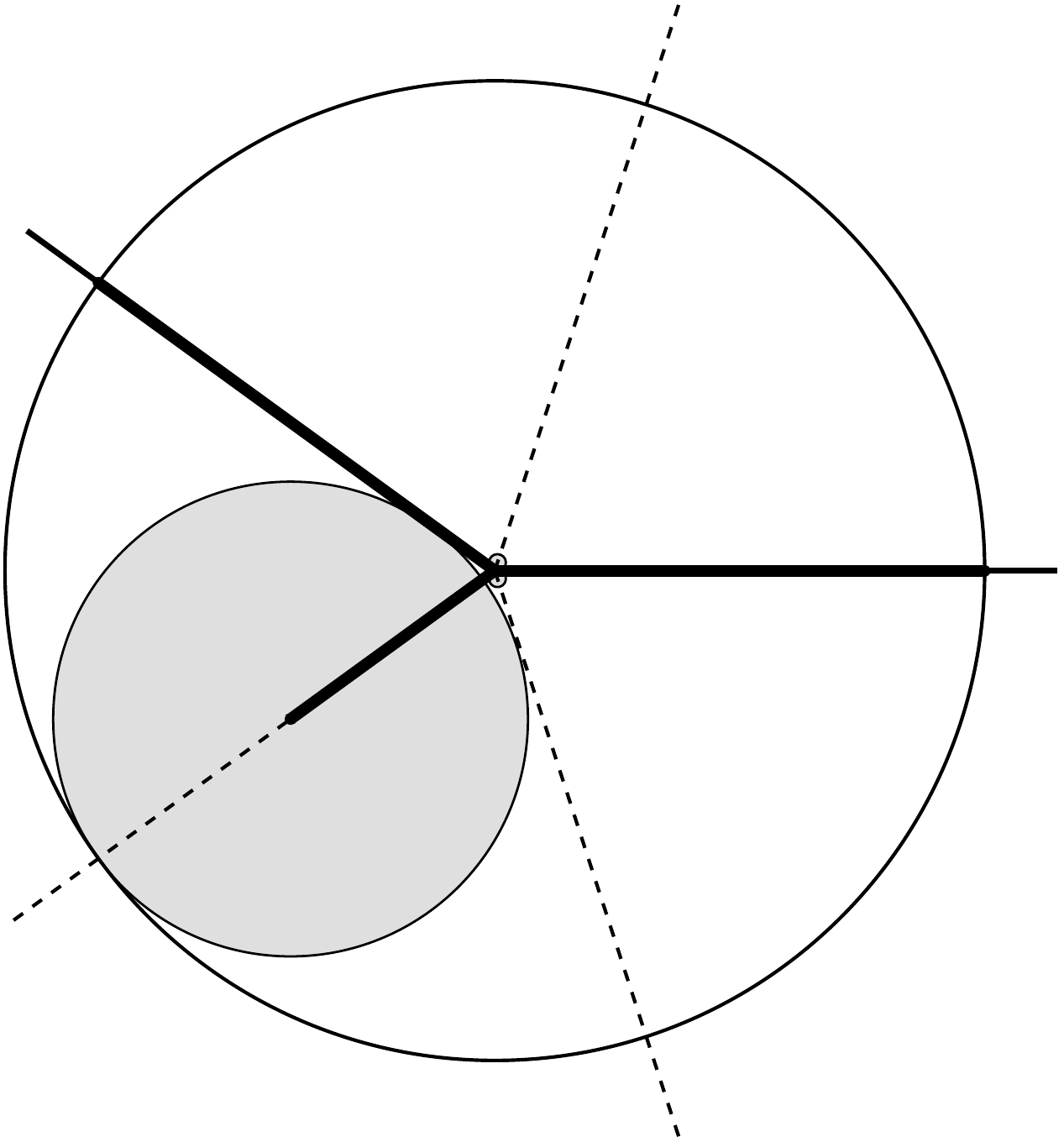}
  \end{center}
  \caption{Three balloons with radius $\varepsilon,\varepsilon,1-2\varepsilon$ and 5 spokes. Separating the unused spokes by an angle   $\ge2\pi/3$ yields a covering disk  with radius  $\alpha(2\pi/5)=\bound$ 
  when $\varepsilon$ approaches zero.}
  \label{fig:limitingcase}
\end{figure}

As a final remark we point out  that Theorem~\ref{thm:general} can be generalized such that it holds for one or two free spokes, while guaranteeing that the whole balloon drawing can be covered with a disk of radius $2$. However, as depicted in Figure~\ref{fig:limitingcase}, the slightly worse bound of $\bound$ cannot be avoided if one has to guarantee that the smaller angle between the two unused spokes is at least  $2\pi/3$. This requirement is however necessary to apply Lemma~2.3 of Duncan~\etal\cite[full version]{DEGKN10}.

%
%
%

\bibliographystyle{abbrv}
\bibliography{balloon}

\clearpage
\appendix

\end{document}